\documentclass[
    aps,
    prb,
    reprint,
    superscriptaddress,
    amsmath,
    amssymb
]{revtex4-2}
\usepackage{graphicx}
\usepackage{placeins}
\usepackage{color}
\usepackage{dcolumn}
\usepackage{latexsym}

\usepackage[normalem]{ulem}
\usepackage{hyperref,amssymb}
\usepackage{url}
\usepackage{color,soul}
\usepackage{graphicx}
\usepackage{verbatim}
\usepackage{multirow}
\usepackage{amsmath}
\newcommand{\beq}{\begin{eqnarray}}
\newcommand{\eeq}{\end{eqnarray}}
\usepackage{mathrsfs}
\usepackage{float,soul}
\usepackage[dvipsnames]{xcolor}
\usepackage{mathtools}
\usepackage{slashed}
\usepackage{physics}
\usepackage{graphicx}
\usepackage{epstopdf}
\usepackage{subfigure}
\usepackage[font=small]{caption}
\usepackage{hyperref}
\usepackage{bbold}
\usepackage{wasysym}
\usepackage{feynmp}
\usepackage{hyperref}
\hypersetup{colorlinks}
\usepackage{ragged2e}
\usepackage{siunitx}
\usepackage[utf8]{inputenc}
\usepackage{bm}

\begin{document}

\title{Interplay of disorder and phonons in two-dimensional transport}
\author{Aryaman Babbar}
\author{Sankar Das Sarma}
\affiliation{Condensed Matter Theory Center and Joint Quantum Institute, Department of Physics, University of Maryland, College Park, Maryland 20742, USA}

\begin{abstract}
We calculate the temperature- and density-dependent resistivity of bilayer MoSe$_2$ in the presence of charged impurities, short-range defects, and longitudinal acoustic phonons, using the Boltzmann transport theory with screening treated in the random phase approximation, and compare the results with recent transport measurements. The disorder parameters are taken from the experiment itself and the phonon parameters are taken from literature estimates for MoSe$_2$. We find that screened acoustic phonons are a small correction to the disorder-limited resistivity over the entire experimental temperature range, so that the temperature dependence of the measured resistivity on the metallic side is controlled by the temperature-dependent screening of charged disorder and by neutral defects. We also show that a resistivity minimum as a function of temperature can arise on both sides of the metal-insulator crossover through two different mechanisms: on the insulating side from the interplay between activated carriers and resistive scattering, and on the metallic side from the interplay between disorder and phonon scattering. The existence of a resistivity minimum by itself therefore does not identify the phase.
\end{abstract}

\maketitle

\section{Introduction}
\label{sec:intro}

Temperature-dependent transport in two-dimensional (2D) semiconductors has attracted renewed interest because experiments on many different materials report a metal-insulator transition (MIT) as the carrier density is lowered \cite{ge2025visualizingnature, yihuangsds, babbaretal_wigner_and_anderson}. These samples invariably contain charged impurities, whose Coulomb potential is long-ranged, and there are situations where short-ranged defects are also present as scattering centers. The carrier density, of electrons or holes, is continuously tunable by gating while the disorder content stays fixed, which makes these systems natural platforms for studying the MIT and the possibility of Wigner crystallization at low density. The observed MIT is not a transition in the thermodynamic sense: it is a crossover. The operational definition used in transport is that the sign of $d\rho/dT$ at the lowest accessible temperature changes at a density $n = n_{\mathrm{MIT}}$, with $d\rho/dT>0$ (metal) for $n>n_{\mathrm{MIT}}$ and $d\rho/dT<0$ (insulator) for $n<n_{\mathrm{MIT}}$. There is no divergent correlation length and no critical exponent. The precise value of $n_{\mathrm{MIT}}$ is itself somewhat ambiguous, since it depends on the temperature range over which the slope is measured.

\begin{figure}[t]
\includegraphics[width=\columnwidth]{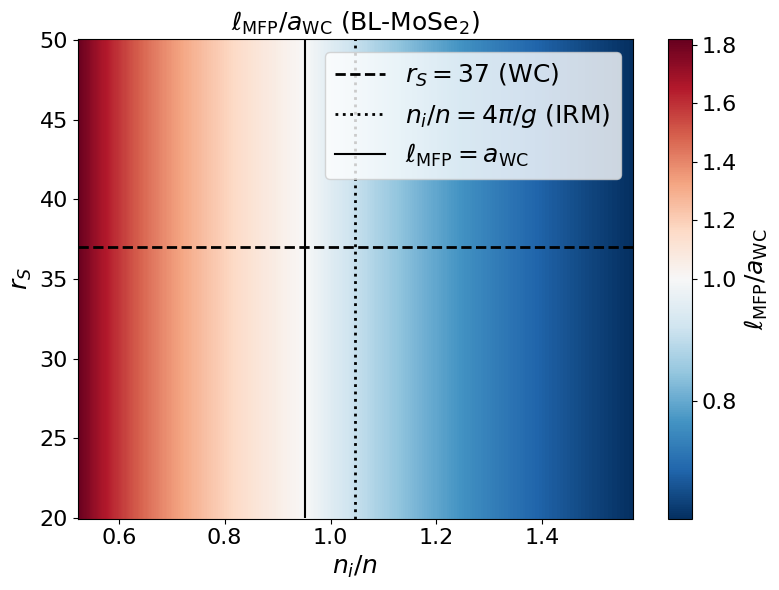}
\caption{Ratio of the mean free path to the Wigner-crystal lattice constant for bilayer MoSe$_2$ as a function of the disorder ratio $n_i/n$ and the interaction parameter $r_S$. The dashed line is $r_S=37$, the dotted line is the Ioffe-Regel-Mott condition $n_i/n = 4\pi/g$, and the solid line is $\ell_{\mathrm{MFP}}=a_{\mathrm{WC}}$.}
\label{fig:mfp}
\end{figure}

A crossover of this kind can arise from several different mechanisms: Wigner crystallization pinned by disorder, carrier trapping and strong (Anderson) localization by disorder, or classical percolation through the disorder landscape. Resistivity-versus-temperature data at a set of densities do not by themselves distinguish among these mechanisms. In practice the observed $n_{\mathrm{MIT}}$ is compared with the theoretical threshold of each candidate mechanism, and the mechanism whose threshold agrees best is claimed to be the cause. Typically, the impurity density in these 2D semiconductors is close to the carrier density at the MIT. Nevertheless, there are experimental claims that because the interaction parameter $r_S = m e^2/(\kappa\hbar^2\sqrt{\pi n})$ at the observed $n_{\mathrm{MIT}}$ is within a factor of order unity of the value $r_S\approx 37$ at which the clean 2D electron liquid is predicted to crystallize \cite{drummond2009phase, 2DEG_GS_QMC_WC2, diffusion_MC_QMC_WC3, vinko_QMC_WC4, flatiron_QMC_WC5}, the observed MIT is a Wigner crystallization. Disorder cannot be left out of this argument. An insulating Wigner crystal must be pinned by disorder to be insulating at all, so the disorder is present by assumption, and given the disorder one must first ask whether the MIT could be a disorder-driven one. Most of the observed transitions occur at $n_{\mathrm{MIT}}\approx n_i$, where $n_i$ is the effective impurity density, which is exactly the density at which a disorder-driven localization crossover is expected. Disorder also shortens the mean free path $\ell_{\mathrm{MFP}}$, and if $\ell_{\mathrm{MFP}}$ is not substantially larger than the Wigner-crystal lattice constant $a_{\mathrm{WC}}$, long-range crystalline order cannot form. So even when $r_S>37$, identifying the insulating state as a Wigner crystal is not justified without further information. A disorder-pinned Wigner crystal can only be observed in a narrow window: the disorder must be strong enough to pin the crystal and make it insulating, but weak enough that $\ell_{\mathrm{MFP}}$ remains larger than $a_{\mathrm{WC}}$, and at the same time the density must be low enough for the interaction to favor crystallization. Figure~\ref{fig:mfp} shows the ratio $\ell_{\mathrm{MFP}}/a_{\mathrm{WC}}$ for bilayer MoSe$_2$ as a function of $n_i/n$ and $r_S$. Here $n_i$ is the effective disorder density defined through the zero-temperature transport scattering rate, $\tau_0^{-1} = \hbar n_i/m$, with $m$ the carrier effective mass, exactly as in Ref.~\cite{babbar20262dtransportinplanemagnetic}; for the sample considered in this paper $n_i$ is set by the measured charged-impurity and defect densities given in Section~\ref{sec:results}.

There is a second, independent reason to doubt a Wigner-crystal interpretation of transport data taken at temperatures of order 1~K. The melting temperature of a 2D Wigner crystal at the relevant densities is of order 100~mK for transition metal dichalcogenides (TMDs) \cite{plasmon, duy_melting_anderson_loc_vs_WC}, so a transport measurement at $\sim 1$~K is already above the melting temperature of any Wigner crystal. Localization observed at such temperatures is therefore likely disorder-induced, since an amorphous Anderson solid ``melts'' at a much higher temperature than a Wigner crystal \cite{duy_melting_anderson_loc_vs_WC}.

Once the measurement temperature exceeds roughly 10~K, disorder alone is no longer the whole story and phonon scattering must be included. Acoustic phonon scattering is suppressed by a power law well below the Bloch-Gr\"uneisen temperature $T_{\mathrm{BG}}$, with the scattering rate behaving as $\tau_{\mathrm{ph}}^{-1}\propto (T/T_{\mathrm{BG}})^{\eta}$ with $\eta\geq 3$ depending on the material and the coupling \cite{kawamura, PhysRevB.77.115449, PhysRevB.86.085307}, and crosses over smoothly to $\tau_{\mathrm{ph}}^{-1}\propto T$ for $T\gg T_{\mathrm{BG}}$. In the high-temperature regime phonons eventually dominate, because the disorder-limited resistivity itself decreases with temperature once $T$ exceeds the Fermi temperature $T_F$ \cite{hwang_metallicity}. This opens the possibility of a nonmonotonic $\rho(T)$ in a metal: at low temperature $\rho$ rises with $T$ because of the thermal weakening of screening \cite{hwang_charged_impurity, hwang_metallicity}, reaches a maximum, falls when $T>T_F$, and then rises again once phonons take over. A limiting form of this behavior, a plateau in $\rho(T)$ over an extended temperature range, has been observed in GaAs \cite{resistivity_saturation}. Whether the full nonmonotonic form is observable in a given material depends on the ordering of $T_F$ and $T_{\mathrm{BG}}$.

In this paper we calculate the resistivity of bilayer MoSe$_2$ as a function of temperature and carrier density, including charged-impurity scattering, short-range defect scattering, and longitudinal acoustic (LA) phonon scattering through a screened deformation potential, and compare the results with the recent Berkeley transport data \cite{ge2025visualizingnature} on the same system whose MIT we analyzed in Ref.~\cite{babbaretal_wigner_and_anderson}. The disorder parameters are those characterized experimentally in Ref.~\cite{ge2025visualizingnature}, and the phonon parameters are taken from literature estimates for MoSe$_2$~\cite{phonon_parameters}. We then discuss the resistivity minimum that appears in $\rho(T)$ on both sides of the crossover, and show that it has a different origin on the two sides. Section~\ref{sec:theory} gives the theory, Section~\ref{sec:results} the results and the comparison with experiment, Section~\ref{sec:valley} the resistivity minima, and Section~\ref{sec:conclusion} the conclusion, including the approximations made and their limitations.

\section{Theory}
\label{sec:theory}

We consider a 2D electron system of carrier density $n$ with parabolic dispersion $\varepsilon_{\bm k} = \hbar^2 k^2/(2m)$, effective mass $m$, and combined spin and valley degeneracy $g$ \cite{yihuangsds}. At temperature $T$ the chemical potential is fixed by the density,
\begin{equation}
\mu = k_B T \ln\!\left[\exp\!\left(\frac{2\pi\hbar^2 n}{g m k_B T}\right)-1\right],
\label{eq:mu}
\end{equation}
so that $\mu\to E_F = 2\pi\hbar^2 n/(gm)$ for $T\ll T_F = E_F/k_B$ and $\mu\to k_B T\ln(T_F/T)$ for $T\gg T_F$.

Since we calculate DC transport, only the static polarizability enters. The finite-temperature static polarizability is \cite{bruus_flensberg_2004, hwang_charged_impurity}
\begin{equation}
\Pi(q,T) = g\!\int\!\frac{d^2k}{(2\pi)^2}\,\frac{n_{\bm k+\bm q}-n_{\bm k}}{\varepsilon_{\bm k}-\varepsilon_{\bm k+\bm q}},
\label{eq:pi_def}
\end{equation}
with $n_{\bm k} = [e^{\beta(\varepsilon_{\bm k}-\mu)}+1]^{-1}$ and $\beta = 1/k_BT$. Writing the Fermi function as a thermal average of step functions,
\begin{equation}
n_{\bm k} = \int_0^\infty d\mu'\,\frac{\Theta(\mu'-\varepsilon_{\bm k})}{4k_BT\cosh^2\!\big[(\mu-\mu')/2k_BT\big]},
\end{equation}
gives the finite-temperature polarizability as a thermal average of the zero-temperature one,
\begin{equation}
\Pi(q,T) = \int_0^\infty d\mu'\,\frac{\Pi_0(q,\mu')}{4k_BT\cosh^2\!\big[(\mu-\mu')/2k_BT\big]},
\label{eq:pi_T}
\end{equation}
where
\begin{equation}
\Pi_0(q,\mu) = \frac{gm}{2\pi\hbar^2}\left[1-\sqrt{1-\frac{8m\mu}{\hbar^2q^2}}\;\Theta\!\left(q^2-\frac{8m\mu}{\hbar^2}\right)\right]
\label{eq:pi_0}
\end{equation}
is the standard 2D static polarizability at $T=0$. The RPA dielectric function is
\begin{equation}
\epsilon(q,T) = 1+\frac{2\pi e^2}{\kappa q}\,\Pi(q,T),
\label{eq:eps}
\end{equation}
with $\kappa$ the background dielectric constant.

The transport scattering rate of an electron of wavevector $k$ due to a random layer of charged impurities, of 2D density $n_c$, located a distance $z$ from the electron plane, is \cite{yihuangsds}
\begin{equation}
\frac{1}{\tau_c(k)} = \frac{4 m n_c}{\pi\hbar^3}\int_0^{\pi/2}\!d\theta\,\sin^2\theta\left[\frac{2\pi e^2}{\kappa q}\frac{e^{-qz}}{\epsilon(q,T)}\right]^2_{q=2k\sin\theta}.
\label{eq:tau_c}
\end{equation}
For the sample considered here the charged impurities are in the plane, $z=0$. Short-range defects of density $n_d$ scatter with a constant matrix element $V_0$ between plane-wave states, and their transport rate is given by the same expression with $2\pi e^2 e^{-qz}/(\epsilon(q,T) \kappa q)$ replaced by $V_0$; we take the defect parameters from Ref.~\cite{babbaretal_wigner_and_anderson}, where this calculation is described in detail.

For LA phonons we follow Refs.~\cite{kawamura, PhysRevB.77.115449, PhysRevB.86.085307}. The electrons couple to the phonons through a deformation potential of strength $D$. The transport scattering rate at energy $\varepsilon_{\bm k}$ is
\begin{equation}
\frac{1}{\tau_{\mathrm{ph}}(\varepsilon_{\bm k})} = \sum_{\bm k'}(1-\cos\theta_{\bm k\bm k'})\,W_{\bm k\bm k'}\,\frac{1-n(\varepsilon_{\bm k'})}{1-n(\varepsilon_{\bm k})},
\label{eq:tau_ph}
\end{equation}
where the transition rate, from the Born approximation, is

\begin{equation}
\begin{aligned}
W_{{\bm k},{\bm k'}}
&=
\frac{2\pi}{\hbar}
\frac{D^2 \hbar q}
{2A\rho_m v\,\epsilon^2(q,T)}
\\
&\quad\times
\Bigg[
N_q\,
\delta\!\left(
\varepsilon_{{\bm k'}}
-\varepsilon_{{\bm k}}
-\hbar vq
\right)
\\
&\qquad+
\left(N_q+1\right)
\delta\!\left(
\varepsilon_{{\bm k'}}
-\varepsilon_{{\bm k}}
+\hbar vq
\right)
\Bigg],
\end{aligned}
\label{eq:W}
\end{equation}

with $q=|\bm k-\bm k'|$, $N_q = [e^{\beta\hbar vq}-1]^{-1}$ the Bose occupation, $v$ the LA sound velocity, $\rho_m$ the 2D mass density of the bilayer, and $A$ the area. The two terms are phonon absorption and emission. The factor $\epsilon^{-2}(q,T)$ is the RPA screening of the deformation-potential coupling by the electrons themselves. For acoustic phonons the relevant wavevectors are $q\leq 2k_F$ and the phonon energy $\hbar vq$ is at most comparable to $E_F$, so static screening of the electron-phonon interaction is the appropriate and standard treatment \cite{kawamura, PhysRevB.86.085307}; the unscreened rate, obtained by setting $\epsilon=1$, overestimates phonon scattering, and we return to this point once in Section~\ref{sec:results}. Equation~\eqref{eq:tau_ph} is evaluated numerically: the delta functions fix $|\bm k'|$ as a function of the angle $\theta_{\bm k\bm k'}$ for the absorption and emission branches separately, and the remaining angular integral is done numerically.

The scattering rates from different mechanisms add at fixed energy, $\tau_{\mathrm{tot}}^{-1}(\varepsilon)=\tau_c^{-1}(\varepsilon)+\tau_d^{-1}(\varepsilon)+\tau_{\mathrm{ph}}^{-1}(\varepsilon)$, and the energy-averaged transport time at temperature $T$ is \cite{yihuangsds}
\begin{equation}
\langle\tau\rangle_T = \frac{\int d\varepsilon\,\varepsilon\,\big(-\partial n/\partial\varepsilon\big)\,\tau_{\mathrm{tot}}(\varepsilon)}{\int d\varepsilon\,\varepsilon\,\big(-\partial n/\partial\varepsilon\big)},
\label{eq:tau_avg}
\end{equation}
where we have used the constant 2D density of states. The resistivity is $\rho(T) = m/(ne^2\langle\tau\rangle_T)$.

The Ioffe-Regel-Mott (IRM) criterion for strong localization, $k_F\ell_{\mathrm{MFP}}\sim 1$ at $T=0$ \cite{IoffeRegel1960, Mott1974, MottDavis1979}, translates in 2D into a critical sheet resistivity
\begin{equation}
\rho_{\mathrm{IRM}}\sim\frac{2}{g}\frac{h}{e^2},
\label{eq:irm}
\end{equation}
which for $g=12$ is $\rho_{\mathrm{IRM}}\approx 4.3$~k$\Omega$. We use this value to mark the boundary of validity of the Boltzmann theory in the figures below.

\section{Results}
\label{sec:results}

The parameters of the bilayer MoSe$_2$ sample are known from the experiment \cite{ge2025visualizingnature} and from our earlier analysis of its MIT \cite{babbaretal_wigner_and_anderson}: in-plane charged-impurity density $n_c = 1.6\times 10^{11}$~cm$^{-2}$, short-range defect density $n_d = 3.5\times 10^{11}$~cm$^{-2}$, effective mass $m=0.54\,m_e$, dielectric constant $\kappa = 3.73$ in the metallic regime, degeneracy $g=12$, and defect strength $(m|V_0|/\hbar^2)^2 = 2$. The transport sample is known to be similar to the STM sample on which these parameters were measured \cite{WangKimPrivateComm}. We use these values as they are, with no adjustment to the transport data.

The phonon parameters are $D\simeq 3.5$~eV and $v\simeq 4.1\times 10^5$~cm/s, which are representative of LA phonons in MoSe$_2$ \cite{phonon_parameters}, and the 2D mass density of the bilayer, $\rho_m\simeq 9.7\times 10^{-7}$~g~cm$^{-2}$, obtained from the bulk density of MoSe$_2$ (6.9~g~cm$^{-3}$) and a monolayer thickness of 0.7~nm \cite{Awais2020MoSe2, Park2018MoSe2Dielectric}. We emphasize that nothing in the phonon calculation is adjusted to the transport data; $D$, $v$ and $\rho_m$ are fixed by the material.

Two temperature scales control the physics: the Fermi temperature $T_F = 2\pi\hbar^2 n/(g m k_B)$ and the Bloch-Gr\"uneisen temperature $T_{\mathrm{BG}} = 2\hbar v k_F/k_B$ with $k_F = \sqrt{4\pi n/g}$. Because of the large degeneracy $g=12$ both scales are low in bilayer MoSe$_2$: at $n=10^{12}$~cm$^{-2}$, $T_F\approx 9$~K and $T_{\mathrm{BG}}\approx 6.5$~K, and at $n=6.5\times 10^{12}$~cm$^{-2}$, $T_F\approx 55$~K and $T_{\mathrm{BG}}\approx 16$~K. So over the experimental window, $T\approx 2$--$100$~K, the system passes from degenerate to nondegenerate at every measured density, and the phonons are in or above the Bloch-Gr\"uneisen crossover for most of the range. There is no regime $T_F\ll T\ll T_{\mathrm{BG}}$ in this material.

The Boltzmann theory is valid only where the RPA-screened perturbative treatment of disorder is valid, that is, in the metallic regime. It becomes progressively worse as $n$ approaches $n_{\mathrm{MIT}}$ from above and does not apply for $n<n_{\mathrm{MIT}}$. In Ref.~\cite{babbaretal_wigner_and_anderson} we argued that the MIT in this sample is a disorder-driven crossover whose density is set by the IRM criterion. Accordingly, we show the calculated resistivity only for densities on the metallic side, we draw the IRM line, \eqref{eq:irm}, on every plot, and we do not expect our curves to be quantitatively meaningful above that line.

\begin{figure*}[t]
    \centering
    \includegraphics[width=\textwidth]{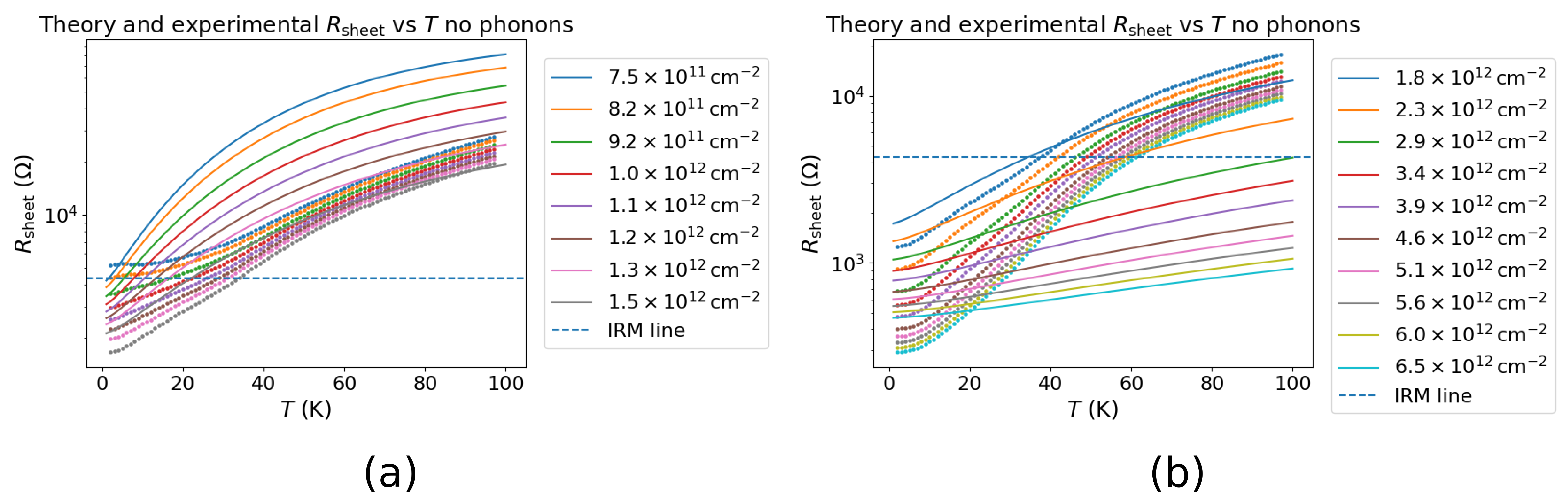}
    \caption{Resistivity versus temperature with charged-impurity and defect scattering only (no phonons), for (a) the lower and (b) the higher carrier densities of the experiment. Solid curves: Boltzmann theory. Dotted curves: experimental data \cite{ge2025visualizingnature}. The dashed horizontal line is the IRM resistivity of \eqref{eq:irm}.}
    \label{fig:noph}
\end{figure*}

\begin{figure*}[t]
    \centering
    \includegraphics[width=\textwidth]{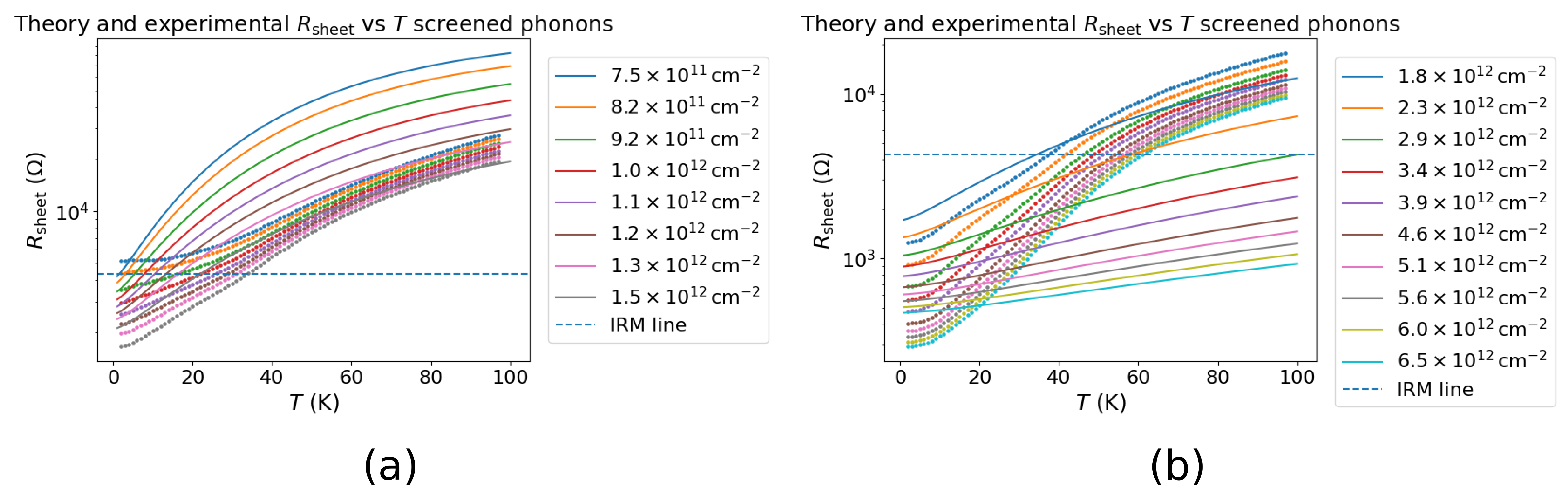}
    \caption{Same as Figure~\ref{fig:noph} with screened LA phonon scattering added, for (a) the lower and (b) the higher carrier densities.}
\label{fig:scr}
\end{figure*}

\begin{figure*}[t]
\centering
    \includegraphics[width=\textwidth]{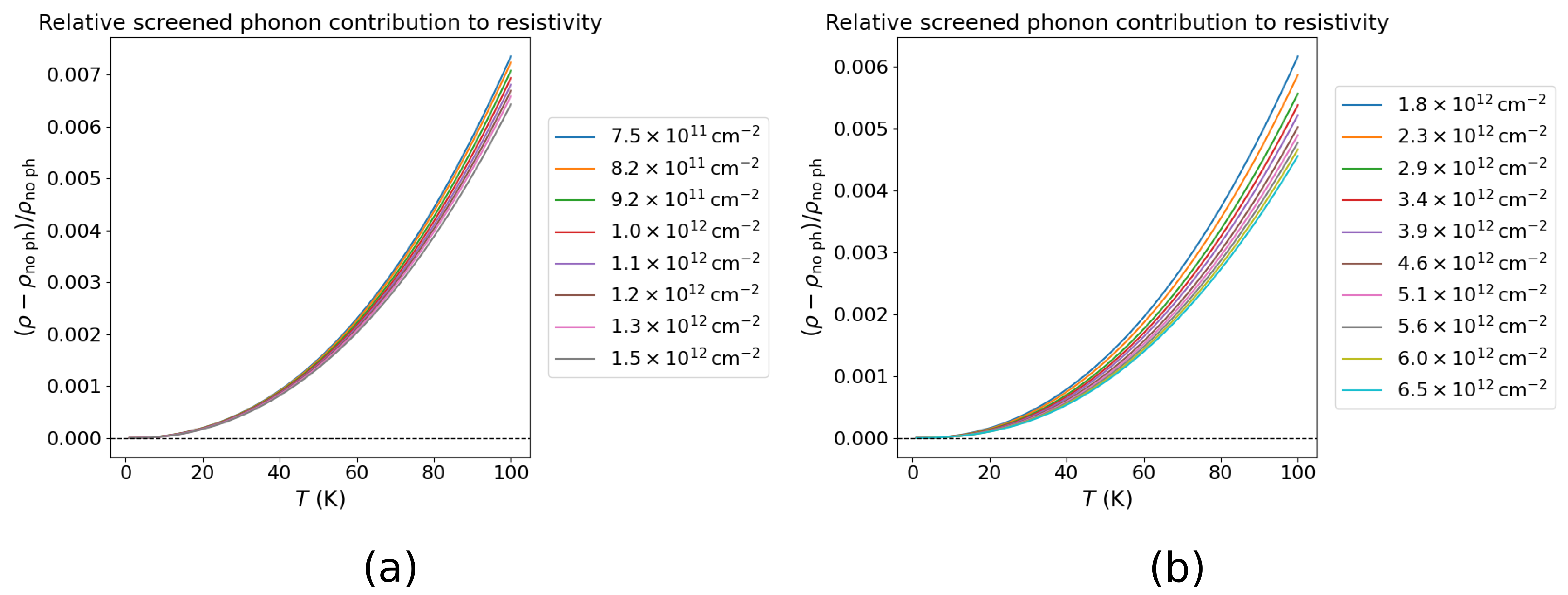}
    \caption{Relative phonon contribution $R(T)$ of \eqref{eq:R} for screened LA phonons, for (a) the lower and (b) the higher carrier densities.}
\label{fig:R}
\end{figure*}

Figure~\ref{fig:noph} shows the disorder-only resistivity, with charged impurities and defects but no phonons, compared with the experimental data at each measured density. Figure~\ref{fig:scr} shows the same comparison with screened LA phonons added. The two sets of theoretical curves are nearly indistinguishable. This is the main quantitative result of the paper: in bilayer MoSe$_2$, over the entire measured temperature range, the temperature dependence of the metallic resistivity comes from the temperature dependence of the screened disorder scattering, and the phonons are a small correction. To quantify the phonon contribution we define
\begin{equation}
R(T) = \frac{\rho(T)-\rho_{\mathrm{no\,ph}}(T)}{\rho_{\mathrm{no\,ph}}(T)},
\label{eq:R}
\end{equation}
where $\rho$ includes phonons and $\rho_{\mathrm{no\,ph}}$ does not. Figure~\ref{fig:R} shows $R(T)$ for screened phonons at each density: it grows monotonically with temperature, is below one percent for $T\lesssim 100$~K. As an aside, we note that the unscreened phonons, obtained by setting $\epsilon=1$ in \eqref{eq:W}, give a contribution roughly 100 times larger, with $R$ exceeding unity at the higher densities at 100~K. The comparison demonstrates the central role of screening in 2D phonon-limited transport; the unscreened result is not physical for this system and we do not consider it further.

The agreement between theory and experiment in Figs.~\ref{fig:noph} and \ref{fig:scr} is best at the higher densities and lower temperatures, where the RPA-Boltzmann theory is most reliable, and degrades at lower densities and higher temperatures, where it is expected to degrade. We stress that no parameter has been adjusted: the disorder parameters are the measured ones, the phonon coupling is the material value, and the curves are predictions, not fits. The theory reproduces the magnitude of the resistivity and its monotonic rise with temperature at every metallic density without any free parameter, which we consider strong evidence that the metallic-side transport in this sample is disorder-limited transport with temperature-dependent screening, as in Si and GaAs 2D systems \cite{hwang_charged_impurity, hwang_metallicity, transport_in_2d_graphene}.

The temperature dependence in the figures can be understood analytically. For $T\ll T_F$ the screened charged-impurity resistivity rises linearly with temperature \cite{hwang_charged_impurity, hwang_metallicity},
\begin{equation}
\rho(T) = \rho_0\left[1+\frac{2x}{1+x}\frac{T}{T_F}+O\!\left(\frac{T}{T_F}\right)^{3/2}\right],
\label{eq:lowT}
\end{equation}
where $\rho_0=\rho(T=0)$ and $x = q_{\mathrm{TF}}/2k_F$ with $q_{\mathrm{TF}}$ the Thomas-Fermi wavevector. For $T\gg T_F$ the same resistivity falls as \cite{hwang_metallicity}
\begin{equation}
\rho(T) = \rho_0\,x^2\,\frac{T_F}{T}\left[1+O\!\left(\frac{T_F}{T}\right)^{3/2}\right],
\label{eq:highT}
\end{equation}
because the nondegenerate electrons screen less effectively and scatter less. For $T\gg T_{\mathrm{BG}}$ the phonon-limited resistivity is linear in $T$ and, in terms of $\lambda = D^2 m/\left(2 \pi \hbar^2 \rho_m v^2\right)$, \cite{kawamura, phonon_mobility_kaasbjerg}

\begin{equation}
\rho_{\mathrm{ph}}(T)\simeq \frac{m}{ne^2}\,\frac{2\pi\lambda\,k_BT}{\hbar}.
\label{eq:rhoph}
\end{equation}
Equation~\eqref{eq:rhoph} is the unscreened result; screening reduces it by the factor $\epsilon^{-2}(2k_F)$ averaged over the Fermi surface, which for bilayer MoSe$_2$ with $g=12$ is a large reduction because $q_{\mathrm{TF}}=g m e^2/(\kappa\hbar^2)$ is large compared with $2k_F$. This is why the screened phonon contribution in Figure~\ref{fig:R} is so small.

The nonmonotonic $\rho(T)$ discussed in the introduction, a rise for $T<T_F$, a fall for $T_F<T<T_{\mathrm{BG}}$, and a phonon-driven rise for $T>T_{\mathrm{BG}}$, requires $T_{\mathrm{BG}}\gg T_F$. In bilayer MoSe$_2$ the two scales are comparable at low density and $T_F>T_{\mathrm{BG}}$ at high density, so the intermediate decreasing regime does not exist and the measured resistivity rises monotonically with temperature at every metallic density, as Figure~\ref{fig:scr} shows. The nonmonotonic behavior would require a material with a larger $g$, a larger $m$, or a larger sound velocity, or some combination of these; we return to this in Section~\ref{sec:valley}.

\section{Resistivity minima on both sides of the crossover}
\label{sec:valley}

Resistivity-versus-temperature curves in 2D semiconductors frequently show a minimum: $\rho$ decreases with increasing $T$ at low temperature, passes through a minimum, and then increases. In this section we show that such a minimum can arise on the insulating side and on the metallic side of the crossover, through two different mechanisms, so that a minimum by itself does not identify which side of the crossover a given density is on.

\begin{figure*}[t]
\centering
    \includegraphics[width=\textwidth]{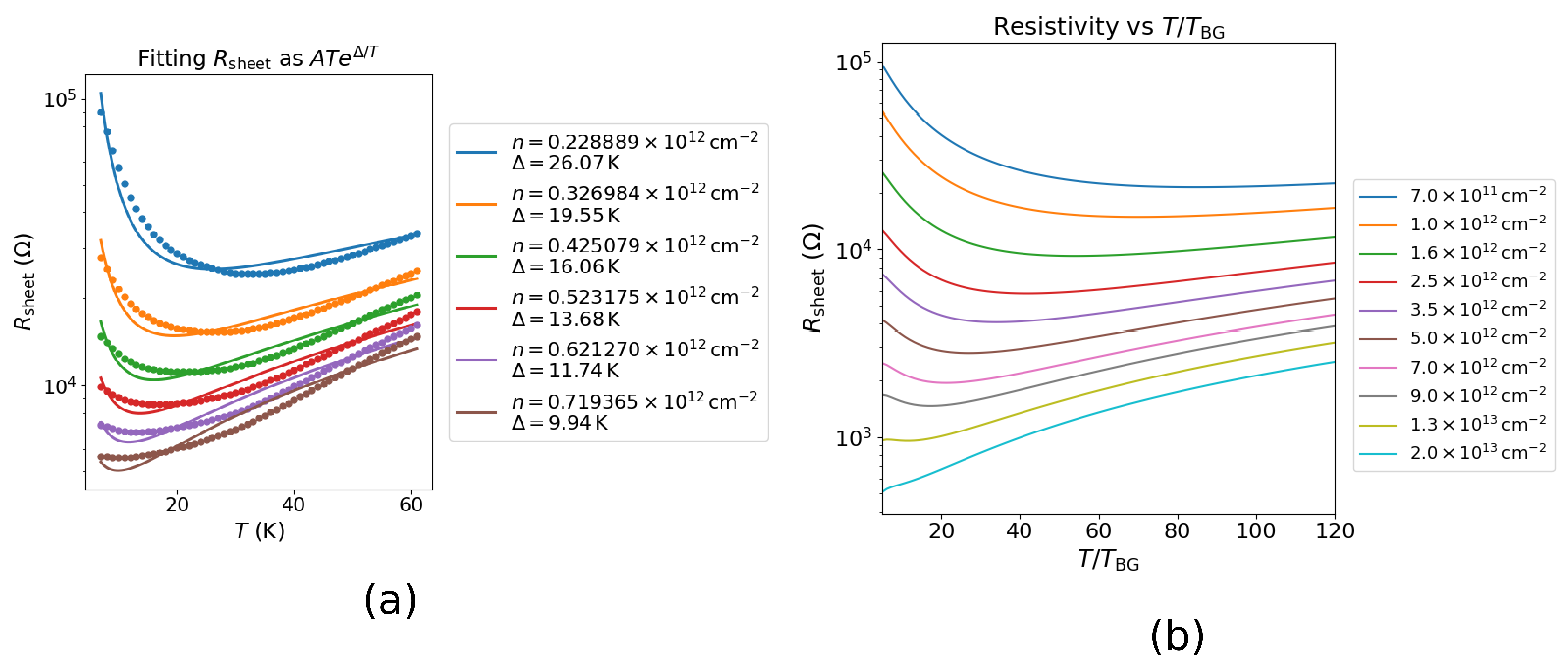}
    \caption{Resistivity minima on the two sides of the crossover. (a) Insulating side: experimental data of Ref.~\cite{ge2025visualizingnature} (dots) at densities below $n_{\mathrm{MIT}}$, with the phenomenological form $\rho(T)=AT\,e^{\Delta/T}$ (solid curves). The minimum arises from the competition between activated carriers at low $T$ and resistive scattering at higher $T$, and the fitted $\Delta$ decreases toward the crossover. (b) Metallic side: calculated Boltzmann resistivity, charged impurities and screened phonons only, plotted against $T/T_{\mathrm{BG}}$ for a hypothetical system with $T_F\ll T_{\mathrm{BG}}$ (bilayer MoSe$_2$ parameters with the sound velocity increased $12$ times). The minimum arises from the crossover between the decreasing disorder-limited resistivity for $T>T_F$ and the increasing phonon-limited resistivity for $T>T_{\mathrm{BG}}$.}
\label{fig:valley}
\end{figure*}

On the insulating side, transport at low temperature is thermally activated, and $\rho$ decreases with $T$ as carriers are excited out of localized states. At higher temperature the number of mobile carriers saturates and their scattering, which increases with temperature, takes over, so $\rho$ increases. A simple phenomenological form that captures the competition is
\begin{equation}
\rho(T) = A\,T\,e^{\Delta/T},
\label{eq:activated}
\end{equation}
where $\Delta$ measures the activation scale and the prefactor linear in $T$ represents the resistive scattering of the activated carriers. Figure~\ref{fig:valley}(a) shows the insulating-side data of Ref.~\cite{ge2025visualizingnature} together with \eqref{eq:activated}. The minimum becomes shallower as the density increases toward $n_{\mathrm{MIT}}$, and the corresponding $\Delta$ decreases, as expected for an activation energy that vanishes at the crossover. We emphasize that \eqref{eq:activated} is phenomenological; we use it only to exhibit the mechanism, not as a theory of the insulating phase.

On the metallic side the minimum has nothing to do with activation. For a material with $T_F\ll T_{\mathrm{BG}}$, the disorder-limited resistivity rises for $T<T_F$ by \eqref{eq:lowT}, falls for $T_F<T<T_{\mathrm{BG}}$ by \eqref{eq:highT} while phonons are still frozen out, and then rises again for $T>T_{\mathrm{BG}}$ by \eqref{eq:rhoph}. The crossover from the disorder-driven decrease to the phonon-driven increase produces a minimum near $T\sim T_{\mathrm{BG}}$. Figure~\ref{fig:valley}(b) shows this for a hypothetical system, with only in-plane charged disorder, constructed from the bilayer MoSe$_2$ parameters with the sound velocity increased by a factor of $12$ ($v = 5 \times 10^6 \text{ cm s}^{-1}$) and the deformation potential constant increased by a factor of $\sim 3$ (D = 10 eV), which puts $T_{\mathrm{BG}}$ well above $T_F$; the calculation includes charged impurities and screened phonons only and is plotted against $T/T_{\mathrm{BG}}$. We stress that this parameter set is not bilayer MoSe$_2$; it is chosen only to make the metallic-side mechanism visible, and the resulting resistivity scale is unphysical. As on the insulating side, the minimum becomes shallower as the density increases, here because the high-temperature decrease of \eqref{eq:highT} is pushed to higher $T_F$ while the phonon rise is unchanged.

The two minima thus look alike on a $\rho(T)$ plot and have opposite physical content: on the insulating side the low-temperature decrease is activation out of localized states, and on the metallic side it is the loss of screening in a nondegenerate metal. The existence of a resistivity minimum therefore cannot be used to assign a given density to the insulating or the metallic side, and in particular cannot be used as evidence for or against Wigner crystallization.

\section{Conclusion}
\label{sec:conclusion}

We have calculated the resistivity of bilayer MoSe$_2$ as a function of temperature and carrier density in the metallic regime, including charged impurities, short-range defects, and screened LA phonons within the RPA-Boltzmann theory, using the disorder parameters measured in the Berkeley experiment \cite{ge2025visualizingnature} and the known electron-phonon coupling parameters of MoSe$_2$, with no adjustable parameter. The theory reproduces the magnitude and the monotonic temperature dependence of the measured metallic resistivity at every density, best at high density and low temperature where the theory is most reliable. Screened phonons are negligible below 100~K, so the measured temperature dependence is the temperature dependence of screened disorder scattering, as in other 2D semiconductors. Unscreened phonons would give a contribution 100 times larger, which demonstrates the decisive role of screening and is not physical for this system. We have also shown that a resistivity minimum arises on both sides of the metal-insulator crossover through different mechanisms, activation on the insulating side and the competition between weakening screening and phonon scattering on the metallic side, so a minimum by itself identifies neither the phase nor the mechanism of the crossover. Together with Ref.~\cite{babbaretal_wigner_and_anderson}, where we argued that the crossover density in this sample is set by the Ioffe-Regel-Mott criterion rather than by Wigner crystallization, the present results support a disorder-driven interpretation of the transport in this system.

The approximations and limitations of the present work are the following. (i) The Boltzmann theory with RPA screening is a weak-disorder theory and is valid only in the metallic regime; it fails as the IRM line is approached and does not apply on the insulating side, where we have used only a phenomenological form. (ii) The electron-phonon coupling is a deformation potential with the material value of $D$ and $v$; piezoelectric coupling, optical phonons, and flexural phonons are neglected, and all of them are expected to be small below 100~K. (iii) The electron-phonon interaction is statically screened in RPA; dynamic screening would matter only for $\hbar vq\gtrsim E_F$, which is marginal at the lowest densities. (iv) The band is taken as parabolic with a single effective mass and a fixed degeneracy $g=12$; the valley degeneracy of the bilayer is itself an assumption, discussed in Ref.~\cite{babbaretal_wigner_and_anderson}. (v) Localization corrections, interaction corrections to the conductivity, and the temperature dependence of the dielectric constant are neglected. (vi) The charged impurities are taken to be in the plane; a finite $z$ would reduce the charged-impurity scattering at large $k_F$. None of these affects the qualitative conclusions, but (i) and (iv) limit the quantitative accuracy at the lowest metallic densities. Future work should extend the theory to the insulating side, where a strong-disorder treatment is required, and should examine other 2D materials with larger sound velocity, where the metallic-side resistivity minimum predicted here should be experimentally accessible.

\begin{acknowledgments}
The authors thank Haleem Kim of UC Berkeley for helpful communications on the experimental data. This work is supported by the Laboratory for Physical Sciences through the Condensed Matter Theory Center. A.B. thanks the Joint Quantum Institute at the University of Maryland for support through a JQI graduate fellowship.
\end{acknowledgments}

\bibliography{references}

\end{document}